\documentclass[preprint,preprintnumbers,superscriptaddress,amsmath]{revtex4}

\usepackage[dvips]{graphics}
\usepackage{epsfig}
\usepackage{dcolumn}
\usepackage{bm}
\usepackage{psfrag}
\usepackage[psamsfonts]{amssymb}
\usepackage{indentfirst}
\usepackage{amssymb}
\usepackage{wrapfig}

\newcommand{\be}{\begin{equation}}
\newcommand{\ee}{\end{equation}}
\newcommand{\ba}{\begin{eqnarray}}
\newcommand{\ea}{\end{eqnarray}}

\begin{document}

\preprint{APS preprint}

\title{Distribution of the Largest Aftershocks
in Branching Models of Triggered Seismicity: Theory of
B{\aa}th's law}

\author{A. Saichev}
\affiliation{Mathematical Department, Nizhny Novgorod
State University, Gagarin prosp. 23, Nizhny Novgorod,
603950, Russia} \affiliation{Institute of Geophysics and
Planetary Physics, University of California, Los Angeles,
CA 90095}

\author{D. Sornette}
\affiliation{Institute of Geophysics and Planetary
Physics and Department of Earth and Space Sciences,
University of California, Los Angeles, CA 90095}
\affiliation{Laboratoire de Physique de la Mati\`ere
Condens\'ee, CNRS UMR 6622 and Universit\'e de
Nice-Sophia Antipolis, 06108 Nice Cedex 2, France}
\email{sornette@moho.ess.ucla.edu}

\date{\today}

\begin{abstract}

Using the ETAS branching model of triggered seismicity,
we apply the formalism of generating probability functions
to calculate exactly the average difference between the
magnitude of a mainshock and the magnitude of its largest
aftershock over all generations. This average magnitude
difference is found empirically to be independent of the
mainshock magnitude and equal to $1.2$, a universal
behavior known as B{\aa}th's law. Our theory shows that
B{\aa}th's law holds only sufficiently close to the
critical regime of the ETAS branching process. Allowing
for error bars $\pm 0.1$ for B\aa th's constant value
around $1.2$, our exact analytical treatment of B\aa th's
law provides new constraints on the productivity exponent
$\alpha$ and the branching ratio $n$: $0.9 \lesssim \alpha \leq 1$
and $0.8 \lesssim n \leq 1$. We
propose a novel method for measuring $\alpha$ based on
the predicted renormalization of the Gutenberg-Richter
distribution of the magnitudes of the largest aftershock.
We also introduce the ``second
B{\aa}th's law for foreshocks:  the probability that a main earthquake turns
out to be the foreshock does not depend on its magnitude $\rho$.

\end{abstract}


\maketitle

\section{Introduction}

This paper is part of our continuing effort to develop a
complete theory of seismicity within models of triggered
seismicity, which allows one to make quantitative
predictions of observables that can be compared with
empirical data
\cite{SS99,HS02,Forexp,Saichevetal04,SaichSorl04,SaiSorpdf}.
We study the general branching process, called the
Epidemic-Type Aftershock Sequence (ETAS) model of
triggered seismicity, introduced by Ogata in the present
form \cite{Ogata} and by Kagan and Knopoff in a slightly
different form \cite{KK81} and whose main statistical
properties are reviewed in \cite{HS02}. The ETAS model
belongs to a general class of branching processes
\cite{Athreya,Sankaranarayanan}, and has in addition the
property that the variance of the number of earthquake
progenies triggered in direct lineage from a given mother
earthquake is mathematically infinite. This model has
been shown to constitute a powerful null hypothesis to
test against other models \cite{Ogata}. The advantage of
the ETAS model is its conceptual simplicity based on
three independent well-found empirical laws
(Gutenberg-Richter distribution of earthquake magnitudes,
Omori law of aftershocks and productivity law) and its
power of explanation of other empirical observations (see
for instance \cite{Forexp} and references therein).

Here, we develop a theoretical formulation based on
generating probability functions (GPF) to construct the
distribution of magnitudes of the largest triggered event
(largest aftershock) within the cascade comprising all
triggered events of a given source earthquake. This
allows us to derive the empirical B{\aa}th's law
\cite{Bath,HSbath}, which states that the average
difference in magnitude between a mainshock and its
largest aftershock is $1.2$ regardless of the mainshock
magnitude, within a completely consistent theory taking
into account all generations of triggered events. Our
present results significantly improve on the numerical
results of \cite{HSbath} by demonstrating the essential
roles played by the cascade of triggered events and the
proximity to criticality in order to obtain B{\aa}th's
law and by providing new improved constraints on the key
parameters of the ETAS model. In addition, we extend
B{\aa}th's law which is a statement on the average
magnitude difference between the mainshock and its
largest aftershock by giving the full distribution. Our
theoretical framework also allows us to calculate
precisely the probability that the largest aftershock
turns out to be larger than its source, a situation which
is usually interpreted as the source and all events
before the largest aftershock being its foreshocks, the
largest aftershock being re-interpreted as the mainshock
of the seismic series.

The paper is organized as follow. The next section 2
recalls the definition of the branching model of
triggered seismicity. Section 3 presents the generating
probability function (GPF) and results on the statistics
of the largest aftershock among aftershocks of the first
generation. The GPF for first-generation aftershocks is
generalized to aftershocks of all generations in section
4. This allows us to predict that the distribution of
magnitudes of the largest aftershock over all aftershock
generations is renormalized in the critical regime. This
renormalization provides a novel way to calibrate the
productivity parameter. Section 5 puts together previous
results to calculate the average difference in magnitude
between the mainshock and its largest aftershock over all
generation. In the critical regime, B{\aa}th's law is
shown to hold. The value of the average difference in
magnitude allows us to offer new improved constraints on
the two key parameters of the ETAS model, the critical
branching ratio $n$ and the productivity exponent
$\alpha$. Section 6 concludes.

\section{The Epidemic-Type Aftershock Sequence (ETAS) branching model
of earthquakes}

Consider an earthquake of magnitude $\rho$, which we
refer to as a mainshock to mean that we are interested in
the earthquakes that it triggers (aftershocks). According
to the ETAS model, it generates a random number
$R_\rho^1$ of first generation aftershocks, which has
Poissonian statistics, \be P_\rho(R_\rho^1)=
e^{-\kappa\mu}
\frac{(\mu\kappa)^{R_\rho^1}}{(R_\rho^1)!}~, \label{mgm}
\ee characterized by the conditional average number \be
N_\rho=\kappa\mu(\rho) , \quad \mu(\rho)=
10^{\alpha(\rho-m_0)} ~.  \label{1} \ee Here $m_0$ is the
minimum magnitude of earthquake capable of triggering
other earthquakes, and $\kappa$ is a constant. The
expression (\ref{1}) for $\mu(\rho)$ is chosen in such a
way that it reproduces the empirical dependence of the
average number of aftershocks triggered directly by an
earthquake of magnitude $m$ (see
\cite{alpha,HelmKaganJackson04} and references therein).
Expression (\ref{1}) gives the so-called productivity law
of a given mother as a function of its magnitude $\rho$.

The ETAS model requires the specification of the
Gutenberg-Richter (GR) density distribution of earthquake
magnitudes \be p(m) = b ~\ln (10)~ 10^{-b (m-m_0)}~,~~~~m
\geqslant m_0~, \label{GR} \ee such that $\int_m^{\infty}
p(x) dx$ gives the probability that an earthquake has a
magnitude equal to or larger than $m$. This magnitude
distribution $p(m)$ is assumed to be independent of the
magnitude of the triggering earthquake, i.e., a large
earthquake can be triggered by a smaller one
\cite{Forexp}. The cumulative ($\mathcal{P}(m)$) and
complementary cumulative ($\mathcal{Q}(m)$) distributions
corresponding to (\ref{GR}) are \be
\mathcal{P}(m)=1-\mathcal{Q}(m), \quad \mathcal{Q}(m)=
10^{-b(m-m_0)}=[\mu(m)]^{-\gamma}~ , \quad
\gamma=b/\alpha ~. \label{2} \ee

The ETAS model is defined by the conditional Poisson
intensity given the average rate of seismicity at time
$t$ and position ${\bf r}$ conditioned on all past
earthquakes \be \lambda(t, {\bf r})= s(t, {\bf r}) +
\sum_{i~|~t_i\leq t} \mu(m_i)~ \Psi(t-t_i)~\phi({\bf
r}-{\bf r}_i)~, \label{mgmkr} \ee where $s(t, {\bf r})$
is the average Poisson rate of spontaneous earthquake
sources (``immigrants'' in the language of epidemic
branching processes) at position ${\bf r}$ and at time
$t$. The sum is over all past earthquakes: each
earthquake is characterized by its occurrence time $t_i$,
its magnitude $m_i$ and its location ${\bf r}_i$ in the
catalog. The two kernels $\Psi(t-t_i)$ and $\phi({\bf
r}-{\bf r}_i)$, whose integrals with respect to time and
space respectively are normalized to $1$, describe the
contribution of the earthquake at time $t_i$ and position
${\bf r}_i$ to the seismic intensity at time $t$ in the
future and at position ${\bf r}$.

\section{Statistics of the largest aftershock among aftershocks of
the first generation}

\subsection{Generating probability
function (GPF) of aftershocks of the first generation
triggered by an arbitrary mainshock}

The Poissonian statistics of the aftershocks of first
generation implies that the generating probability
function (GPF) of their numbers reads \be
\Theta_1(z|\rho)=\langle z^{R_\rho^1}\rangle=
\sum_{r=0}^\infty P_\rho(r) z^r=
e^{\kappa\mu(\rho)(z-1)}~ , \ee where we have used
(\ref{mgm}) and the angle brackets correspond to taking
the statistical averaging.

Let $M_1,~ M_2, \dots M_{R^1_\rho}$ be the random
magnitudes of the $R^1_\rho$ aftershocks of first
generation triggered by the mainshock. Let us consider
the statistical average defined by \be
\Theta_1(z,m|\rho)= \langle z^{R^1_\rho}
\prod_{k=1}^{R^1_\rho} H(m-M_k)\rangle ~, \ee where $H$
is the Heaviside function. Using the Poissonian
statistics of the random number $R^1_\rho$ of aftershock
numbers and the Gutenberg-Richter law  for their
magnitudes, we obtain \be \Theta_1(z,m|\rho)=
\sum_{r=0}^\infty P_\rho(r) z^r [\mathcal{P}(m)]^r=
e^{\kappa\mu[z \mathcal{P}(m)-1]}. \label{3} \ee Notice
that $\Theta_1(z,m|\rho)$ can be rewritten as \be
\Theta_1(z,m|\rho)= \langle z^{R^1_\rho(m)} H(m-
M_\rho^1)\rangle ~, \label{4} \ee where $M_\rho^1$ is the
largest magnitude over all aftershocks of the first
generation triggered by the mainshock and $R^1_\rho(m)$
is the number of aftershocks for those realizations of
aftershocks in which no aftershocks of the first
generation exceed the magnitude $m$.

The interest in $\Theta_1(z,m|\rho)$ in (\ref{3},\ref{4})
lies in particular in the fact that, for $z=1$, it
reduces to the probability $P_1(m|\rho)$ that the largest
magnitude among all aftershocks of the first generation
is smaller than $m$: \be
P_1(m|\rho)=\Theta_1(z=1,m|\rho)=
\text{Pr~}\{M_\rho^1<m\}= e^{-\kappa\mu(\rho)
\mathcal{Q}(m)} ~. \label{5} \ee

\subsection{GPF of aftershocks of the first generation triggered by
a spontaneous source}

Consider now some spontaneous source (contributing to the
term $s(t, {\bf r})$ in (\ref{mgmkr})), with magnitude
$M_0$. According to the ETAS model, it triggers its own
aftershocks sequence independently of all other
sequences. Let \be M_0,~ M_1,~ M_2, \dots M_{R^1}
\label{6} \ee be the random sequence of magnitudes
(including $M_0$) of the first generation aftershocks
triggered by the spontaneous source. Then, analogously to
(\ref{3}), the GPF \be \Theta_1(z,m)= \langle z^{R^1(m)}
H(m-M^1)\rangle~ , \ee describing the statistics of the
number $R^1(m)$ of aftershocks of first generation
(triggered by the spontaneous source) and their largest
magnitude $M^1$ among the list (\ref{6}), is equal to \be
\Theta_1(z,m)= F[1-z\mathcal{P}(m)]-[\mu(m)]^{-\gamma}
F(\mu(m)[1-z\mathcal{P}(m)]) ~, \label{7} \ee where \be
F(y)= \gamma \kappa^\gamma y^\gamma \Gamma(-\gamma,\kappa
y) ~. \label{8} \ee For $m\to \infty$, the GPF
$\Theta_1(z,m)$ given by (\ref{7}) reduces to the
standard GPF of the random number $R_1$ of aftershocks of
the first generation triggered by some spontaneous
source, which reads \be \Theta_1(z)=\Theta_1(z,m=\infty)=
F(1-z) ~. \ee

In the following analysis, the function $F(y)$ in
(\ref{8}) plays a crucial role. It is thus useful to
state some of its analytical properties. For
$1<\gamma\lesssim 1.5$ and $y\lesssim 0.2$, it can be
represented rather accurately by \be F(y)\simeq
1-n\,y+\beta\,y^\gamma , \qquad \beta=-
\left(n\frac{\gamma-1}{\gamma}\right)^\gamma
\Gamma(1-\gamma) ~. \label{10} \ee

In the theory of aftershocks branching processes, the
branching ratio, equal to the average number of first
generation aftershocks triggered by some mother
earthquake, plays a fundamental role since it controls
the subcriticality versus supercriticality of the
process. It is defined as \be n=\left.
\frac{d\Theta_1(z)}{dz}\right|_{z=1}= \frac{\kappa
b}{b-\alpha}~ . \ee Thus, for given $b$, $\alpha$ and
$n$, the constant $\kappa$ in relations (\ref{3}) and
(\ref{5}) can be replaced by \be
\kappa=\kappa(\gamma,n)=n
\left(1-\frac{\alpha}{b}\right)\,. \ee

Taking $z=1$ in (\ref{7}) obtains the cumulative
distribution function (CDF) $P_1(m)$ of the largest
magnitude $M^1$ among the sequence (\ref{6}) of
aftershocks including their spontaneous source \be
P_1=F(\mathcal{Q})-\mu^{-\gamma} F(\mu\mathcal{Q})\,. \ee

The GPF $\Theta_1(z,m)$ in (\ref{7}) describes the
statistics of the spontaneous source and its first
generation aftershocks, such that all magnitudes,
including the magnitude of the spontaneous source, are
smaller than $m$. In the following, we will consider the
possibility that the largest aftershock may be larger
than the source, a situation which is known in the
seismological literature as the occurrence of foreshocks
(see \cite{JonesMolnar,HSG03,Forexp} and references
therein). The corresponding GPF, averaged over all
possible foreshock's magnitudes, is denoted as
$\bar{\theta}_1(z,m)$ and is obtained formally from
(\ref{7}) by taking the limit $\mu\to\infty$: \be
\bar{\theta}_1(z,m)= F[1-z\mathcal{P}(m)] ~. \label{9}
\ee

\subsection{Magnitude of the largest aftershock of the first
generation aftershocks}

Before analyzing the conditions under which the empirical
B{\aa}th's law can be obtained from the ETAS model, it is
useful to ask what is its analog when restricting the set
of aftershocks to the first generation of events
triggered by the source. This will provide a reference
point against which to gauge the impact of the multiple
generations of aftershocks on B{\aa}th's law.

We start from expression (\ref{5}) giving the CDF
$P_1(m|\rho)$ of the largest magnitude $M_\rho^1$ among
all aftershocks of the first generation. Substituting
(\ref{1},\ref{2}) in (\ref{5}), we obtain 
\be
P_1(m|\rho)=G[w_0(m-\rho+v_0)] ~, \label{11} 
\ee 
where
\be 
v_0(\rho)=\left(1-\frac{\alpha}{b}\right)(\rho-m_0)+
\frac{1}{b}\log_{10} \left(\frac{b}{n(b-\alpha)}\right)~,
\qquad w_0=b \ln 10~, \label{12} \ee and \be
G(x)=\exp\left(-e^{-x}\right) \label{mgml} 
\ee 
is the
well-known limiting extremal Gumbel  CDF.

It follows from (\ref{11}) in particular that the
probability density function (PDF) of the difference \be
\Delta_\rho^1=\rho-M_\rho^1 \label{13} \ee between the
source (mainshock) magnitude and the magnitude of its
largest aftershock of the first generation is equal to
\be f_1(\delta|\rho)= w_0 ~g[w_0(v_0-\delta)] ~,
\label{14} \ee where \be g(x)=\exp(-x-e^{-x}) \ee is the
PDF associated with the CDF (\ref{mgml}). Note that the
shape and variance of the PDF (\ref{14}) does not depend
on the mainshock magnitude $\rho$. Only its mode
$v_0(\rho)$ (most probable value of the difference
(\ref{13})) depends on $\rho$ and increases linearly with
it according to (\ref{12}).

These results treat all aftershock sequences on the same
footing and in particular include sequences which have
zero aftershocks. In a real data analysis, the
statistical properties of the difference (\ref{13}) are
obtained conditioned on the observation of at least one
aftershock, which requires a modification of the
expressions above. We are interested in modifying the CDF
(\ref{5}) to eliminate the cases where $R_\rho^1=0$. This
corresponds to obtaining the CDF of the largest magnitude
of first generation aftershocks under the condition that
the mainshock triggers at least one aftershock: \be
P_1(m|\rho;1)=1-Q_1(m|\rho;1) , \quad
Q_1(m|\rho;1)=\frac{1-e^{-\kappa\mu(\rho)
\mathcal{Q}(m)}}{1-e^{-\kappa\mu(\rho)}} ~. \label{15}
\ee The corresponding PDF of the difference (\ref{13})
reads \be f_1(\delta|\rho)= w_0 \frac{
g[w_0(v_0-\delta)]}{1-e^{-\kappa\mu(\rho)}} , \qquad
-\infty< \delta <\rho-m_0 ~. \label{16} \ee The
conditional CDF (\ref{15}) differs significantly from the
unconditional one (\ref{5}) only if the probability
$e^{-\kappa\mu(\rho)}$ that there are no aftershocks is
close to $1$. This occurs for small mainshock magnitudes.
How small should the mainshocks be for this difference to
be important? Let us define a magnitude threshold
$\rho_0$ by \be \kappa\mu(\rho)\simeq 2 \qquad
\Rightarrow \qquad \rho_0=m_0+ \frac{1}{\alpha} \log_{10}
\left( \frac{2b}{n(b-\alpha)}\right) ~. \ee For mainshock
magnitudes $\rho>\rho_0$, the conditional CDF (\ref{15})
does not differ significantly from the unconditional one
(\ref{5}). In this case, relations (\ref{14}) and
(\ref{16}) are approximately equal, and the looked for
distribution of the difference (\ref{13}) can be taken to
be the PDF (\ref{14}) where $\delta\in(-\infty,\infty)$.
Thus, for $\rho>\rho_0$, the average of the magnitude
difference (\ref{13}) can be approximated by \be
\Delta_\rho^1 m\equiv \rho-\langle M_\rho^1\rangle\simeq
w_0 \int_{-\infty}^\infty \delta~ g[w_0(v_0-\delta)]
d\delta= v_0(\rho)-\nu/w_0~, \label{17} \ee where
$\nu\simeq 0.5772$ is the Euler constant. Figure 1 shows
the exact average difference $\Delta_\rho^1 m$ calculated
with (\ref{16}), its approximation (\ref{17}) valid for
sufficiently large mainshocks  $\rho>\rho_0$ and the most
probable value $\Delta_\rho^1 m_*=v_0(\rho)$ of the
difference in magnitude between the mainshock magnitude
$\rho$ and its largest aftershock. This figure is typical
of the strong dependence found for all reasonable values
of the parameters and distinguishes this result from the
empirical B{\aa}th's law (which gives a constant value
independent of $\rho$).

\section{Statistics of the largest aftershock among aftershocks of
all generations}

\subsection{GPF of the aftershocks over all generations triggered by
a spontaneous source}

Due to the mutual statistical independence of different
branches of triggered earthquakes in the ETAS model, one
can easily generalize the results for the largest
aftershock of the first generation to derive the
statistical properties of the largest aftershock over all
generations.

Within the ETAS branching model, taking into account all
aftershocks of all generations which are triggered by the
mainshock amounts to replacing in the r.h.s. of equation
(\ref{3}) the PDF $\mathcal{P}(m)$ of the magnitudes of
single aftershocks by the GPF $\Theta(z,m)$ for all
aftershocks triggered by some spontaneous source which
have (together with the source) magnitudes smaller than
$m$. As a result, we obtain the sought GPF of the number
of all aftershocks triggered by the mainshock conditioned
on all magnitudes to be less than $m$ as \be
\Theta(z,m|\rho)=\langle z^{R_\rho(m)}\, H(m-
M_\rho)\rangle= e^{\kappa\mu(\rho)\left[z\Theta(z,m)-1
\right]} ~. \ee Here, $M_\rho$ is the magnitude of the
largest aftershock. In particular, the complementary CDF
of the magnitude of the largest aftershock reads 
\be
Q(m|\rho)= \langle H(M_\rho -m)\rangle=
1-e^{-\kappa\mu(\rho)Q(m)} ~. \label{18} 
\ee

Similarly, the functional equation for the GPF
$\Theta(z,m)$ is obtained by replacing in (\ref{7}) both
$\Theta_1$ and $\mathcal{P}(m)$ by $\Theta(z,m)$: \be
\Theta(z,m)= F(1-z \Theta(z,m))-\mu^{-\gamma}(m)
F[\mu(m)(1-z \Theta(z,m))] ~. \label{19} \ee For $z=1$,
equation (\ref{19}) reduces to an equation for the CDF
$P(m)$ of the magnitude $M$ of the largest event (including
all aftershocks and the source) defined as \be
\Theta(z=1, m) = P(m)= \text{Pr\,}\{M<m\} ~. \ee This
equation reads \be P=F(Q)-\mu^{-\gamma} F(\mu Q) , \qquad
Q(m)=1-P(m) ~. \label{20} \ee Notice that in the limit
$m\to\infty$, Eq.~(\ref{19}) reduces to the well-known
functional equation \be \Theta=\Theta_1(z\Theta) \ee for
the standard  GPF \be \Theta(z, m=\infty)=\Theta(z)=
\langle z^{R}\rangle \ee of the random number $R$ of all
aftershocks triggered by some ancestor, which has been
studied in \cite{Saichevetal04,SaiSorpdf}.

Similarly to the reasoning leading to (\ref{15}), the
conditional probability that the magnitude of the largest
aftershock exceeds $m$, under the condition that the
mainshock triggers at least one aftershock, is \be
Q(m|\rho;1)=\frac{1-e^{-\kappa\mu(\rho)
Q(m)}}{1-e^{-\kappa\mu(\rho)}}~, \qquad Q(m)=1-P(m)~.
\label{21} \ee

It is also of interest to obtain the GPF
$\bar{\Theta}(z,m)$ of the number of aftershocks of all
generations with magnitudes smaller than $m$ which are
triggered by some spontaneous source of arbitrary
magnitude. It is given by replacing in the r.h.s. of
Eq.~(\ref{9}) $\mathcal{P}(m)$ by the GPF $\Theta(z,m)$
given by (\ref{19}) which yields \be \bar{\Theta}(z,m)=
F[1-z\Theta(z,m)]~ . \ee For $z=1$, this gives the
probability $\bar{P}(m)$ that the magnitude of the
largest aftershock triggered by an arbitrary spontaneous
source is smaller than $m$: \be \bar{P}(m)=F[Q(m)] ~.
\label{22} \ee

\subsection{Distribution of the magnitude of the largest aftershock
of a spontaneous earthquake source of arbitrary
magnitude}

All quantities defined above require the knowledge of
$Q(m)$, which has a straightforward statistical meaning:
it is the complementary CDF of the magnitude of the
largest aftershock of a spontaneous source of arbitrary
magnitude (in other words, over all possible source
magnitudes). It is easy to calculate $Q(m)$ by solving
equation (\ref{20}) numerically. We can also use the
algebraic approximation (\ref{10}) of the function $F(y)$
to obtain an explicit and rather precise analytic
expression of $Q(m)$. Indeed, substituting (\ref{10})
into (\ref{20}) obtains \be Q(m)\simeq \frac{1}{(1-n)
[\mu(m)]^\gamma +n\mu(m)} ~. \label{23} \ee This
expression shows that there is a cross-over magnitude
$m_c$, given by \be \mu(m_c)\simeq
\left(\frac{n}{1-n}\right)^{1/(\gamma-1)} \quad
\Rightarrow \quad m_c\simeq m_0+ \frac{1}{b-\alpha}
\log_{10}\left(\frac{n}{1-n}\right)~ , \label{24} \ee
separating two regimes with different power laws for
$Q(m)$. The first regime 
\be Q(m) \simeq \frac{1}{n} 10^{-\alpha (m-m_0)} , 
\qquad m \lesssim m_c~, 
\label{25}
\ee 
corresponds to a complementary CDF decaying slower
than the Gutenberg-Richter law (\ref{GR},\ref{2}), for
$\alpha < b$. In the critical case $n=1$, $m_c=\infty$
and this regime (\ref{25}) holds for any $m>m_0$. The
second regime recovers the Gutenberg-Richter law \be Q(m)
\simeq \frac{1}{1-n} 10^{-b (m-m_0)} , \qquad  m\gtrsim
m_c~. \label{26} \ee Figure 2 shows the logarithm (in
base $10$) of the complementary CDF $Q(m)$ as a function
of $m-m_0$ and the two power law asymptotics (\ref{24})
and (\ref{25}). We have thus shown that the
Gutenberg-Richter law can be renormalized from a bare
exponent $b$ to a smaller exponent $\alpha$ when the
distribution is restricted to the set of largest
aftershocks of spontaneous earthquakes of arbitrary
magnitudes. This renormalization of the $b$-value from
$b$ to $\alpha$ is intrinsically a cascade phenomenon. In
other words, it results from the existence of a cascade
of triggered earthquakes over many generations as shown
in details in \cite{Saichevetal04}. This renormalization
proceeds by a mechanism similar to that of the $p$-value
of the Omori law from a value $1+\theta$ to $1-\theta$
\cite{SS99,HS02}. It is different from the mechanism
leading to an exponent $b-\alpha$ for the asymptotic
branch of the Gutenberg-Richter distribution of all
foreshocks \cite{HSG03}.

This prediction (\ref{25}) offers a novel method for
measuring the key exponent $\alpha$ controlling the
productivity or triggering efficiency of earthquakes as a
function of their magnitude, according to (\ref{1}). What
is needed to implement this new method is a declustering
technique to identify the spontaneous sources and their
largest aftershocks. The statistical declustering
technique of Zhuang et al. \cite{Zhuangetal1,Zhuangetal}
seems to be particularly suitable for this purpose.

Due to the independence between the aftershock sequences
of different sources in the ETAS model, it is
straightforward to obtain the probability distribution of
the largest triggered events among a set of $r$
spontaneous sources. The corresponding complementary CDF,
giving the probability that the largest event triggered
over all generations by $r$ sources is larger than $m$,
is equal to \be \bar{Q}(m|r)=1-F^r[Q(m)] ~. \label{27}
\ee Figure 3 plots $\bar{Q}(m|r)$ as a function of
$m-m_0$, for $r$ running from $1$ to $15$ for
$\alpha=0.8$, $b=1$ and for two values of $n$, $n=0.9$
and $n=1$. It is clear that most of the largest triggered
events in aftershock sequences are very small, simply due
to the interplay of two factors: most random sources are
themselves small and have a small productivity, and the
Gutenberg-Richter distribution makes it much more
probable that all triggered events have small magnitudes.

\subsection{Distribution of the magnitude of the largest aftershock
of a spontaneous earthquake source of fixed magnitude
$\rho$}

Rather than considering arbitrary source magnitudes, it
is interesting to determine the complementary CDF
$Q(m|\rho,r)$ of the magnitude of the largest event
triggered by $r$ spontaneous sources with fixed
magnitudes $\rho_1,~ \rho_2, \dots, \rho_r$. It is
closely approximated by 
\be 
Q(m|\rho,r)\simeq
1-\exp\left[-\kappa Q(m) \sum_{i=1}^r \mu(\rho_i) \right]~. 
\label{28} 
\ee 
In the following, we restrict our
analysis to the case of a single $r=1$ spontaneous source
which fixed magnitude $\rho$. In this case, expression
(\ref{28}) transforms into (\ref{18}).

For $m<m_c$, where the cross-over magnitude $m_c$ is
defined by (\ref{24}), expression (\ref{18}) can be
simplified by replacing the exact $Q(m)$ by the
approximation (\ref{25}), which gives 
\be 
Q(m|\rho)\simeq
1-G\left(w_1(m-\rho+v_1)\right) \qquad m_0<m\lesssim m_c ~. 
\label{29} 
\ee 
where 
\be 
v_1=\frac{1}{\alpha}
\log_{10}\left(\frac{b}{b-\alpha}\right)~, \qquad
w_1=\alpha \ln 10 ~. \label{30} 
\ee
Expression (\ref{29}) can be further simplified into 
\be 
Q(m|\rho)\simeq  {\kappa \over n}
~10^{-\alpha (m - \rho)} = \left( \kappa ~10^{\alpha
(\rho - m_0)}\right) \times \left(
\frac{1}{n}~10^{-\alpha (m - m_0)} \right)~, \label{exjs}
\ee 
in the
tail of $Q(m|\rho)$, i.e., for $\rho - {1 \over \alpha} \log_{10} {n \over \kappa} < m$.
The re-writing of $Q(m|\rho)$ under the form shown by
the last equality in (\ref{exjs}) clarifies its origin:
the first factor $\kappa ~10^{\alpha (\rho - m_0)}$ is
nothing but the productivity law (\ref{1}); the second
factor $\frac{1}{n}~10^{-\alpha (m - m_0)}$ is the
renormalized Gutenberg-Richter law (\ref{25}).

For $m > m_c$, we obtain another approximation for
$Q(m|\rho)$ by replacing in the r.h.s. of equation
(\ref{18}) the complementary CDF $Q(m)$ by the
approximation (\ref{26}), which yields \be
Q(m|\rho)\simeq 1-G\left(w_2[m-\rho+v_2(\rho)]\right)
\qquad m\gtrsim m_c ~, \label{31} \ee where \be v_2=
\left(1-\frac{\alpha}{b}\right) (\rho-m_0)+ \frac{1}{b}
\log_{10}\left(\frac{b(1-n)}{n(b-\alpha)}\right)~ ,
  \qquad w_2=b \ln 10~.
  \label{32}
\ee

The following approximation including both regimes
$m<m_c$, $m>m_c$ and laws (\ref{29},\ref{31}) is obtained
from (\ref{18}) by replacing $Q(m)$ by the approximation
(\ref{23}): 
\be 
Q(m|\rho)\simeq 1-\exp\left(
-\frac{\kappa \mu(\rho)}{(1-n)\mu^\gamma(m)+n
\mu(m)}\right)~ . \label{35} 
\ee 

Figures 4 and 5 present the dependence of the
complementary CDF $Q(m|\rho)$ as a function of the
magnitude of the largest event triggered by a spontaneous
source of fixed magnitude $\rho$, for four different
values of $\rho$. The figures show the exact $Q(m|\rho)$
obtained numerically, its approximation (\ref{35}) (which
is actually undistinguishable from the exact one) and the
universal approximation (\ref{29}). The comparison
between Figure 4 (for $\alpha=0.8$) and Figure 5 (for
$\alpha=0.9$) shows that the approximation (\ref{29})
becomes more and more precise as $\alpha$ becomes closer
to $b$.

As a bonus, we obtain the probability $Q_*(\rho)$ that
the magnitude $m$ of the largest aftershock exceeds the
source magnitude $\rho$, a situation which is usually
classified in seismic catalogs by saying that the largest
triggered event is the mainshock and the spontaneous
source that initiated the sequence and all triggered
events before the largest aftershock are foreshocks.
Indeed, $Q_*(\rho)$ is nothing but 
\be 
Q_*(\rho)\equiv Q(\rho|\rho;1) ~. \label{33} 
\ee 
Interestingly, in the
regime $m_0<\rho < m_c$ (i.e., for $n$ sufficiently
close to $1$ and/or $\alpha$ close to $b$) for which
(\ref{29}) holds, we obtain 
\be 
Q_*(\rho)\simeq
\text{const}=Q_*=1- \exp\left(\frac{\alpha-b}{b}\right)~, 
\label{34} 
\ee 
which is independent of $\rho$ and of the
branching ratio $n$. This approximation is all the
better, the closer $\alpha$ is to $b$.
Figure 6 shows the exact $Q_*(\rho)$ as a function of $\rho$,
which can be compared with the constant (\ref{34}). One can 
observe that, at least for $\alpha = 0.95$,  $Q_*(\rho)$  is actually
quite close to the constant (\ref{34}) over all possible magnitudes
$\rho > m_0$. We propose to call the prediction (\ref{34}) the ``second
B{\aa}th's law for foreshocks:  the probability that a main earthquake turns
out to be the foreshock does not depend on its magnitude $\rho$ (more
generally, the distribution of the difference $\rho-m$ does not
depend on $\rho$).

\section{Derivation of B\aa th's law}

The derivation of the distribution (\ref{21}) and the
approximations (\ref{29}) and (\ref{31}) allow us to
derive B\aa th's law by calculating the statistical
average \be \Delta_\rho m=\rho-\langle M_\rho\rangle~,
\label{36} \ee where $M_\rho$ denotes the magnitude of
the largest aftershock among all events of all
generations triggered by the source of fixed magnitude
$\rho$. Recall that B\aa th's law states that
$\Delta_\rho m$ is independent of $\rho$ and equal to
$1.2$.

Within the ETAS model, the exact value of $\Delta_\rho m$
is obtained as \be \Delta_\rho m=\rho- \int_{m_0}^\infty
Q(m|\rho;1) dm \label{39} \ee where $Q(m|\rho;1)$ is
given by (\ref{21}) and the r.h.s. of (\ref{39})
expresses the fact that the average is performed over
sequences with at least one aftershock.

The two regimes $m<m_c$ giving the asymptotic
(\ref{29}) and $m>m_c$ giving the asymptotic
(\ref{31}) provides two asymptotic expressions for
$\Delta_\rho m$. Indeed, calculating $\langle
M_\rho\rangle$ using the approximation (\ref{29}) and
neglecting the boundary effects (i.e., supposing that
$m\in(-\infty,\infty)$) obtains 
\be 
\Delta_\rho m_1\simeq
v_1-\frac{\nu}{w_1}=B=\frac{1}{\alpha} \left[
\log_{10}\left(\frac{b}{b-\alpha}\right)- \frac{\nu}{\ln
10}\right] ~, \label{37} 
\ee 
which is independent of
$\rho$. Recall that $\nu \simeq 0.5772$ is Euler
constant. In the following, we call $B$ defined in
(\ref{37}) the B\aa th's constant. Note that this regime
$m<m_c$ corresponds to the critical branching regime
of $n$ close to $1$ (for a fixed magnitude $\rho$) and
expresses the full effect of the cascade of triggered
events over all possible generations. It is remarkable
that the theory of the ETAS branching model predicts the
first part of B\aa th's law that the average of the
difference between the magnitude of a mainshock and its
largest aftershock is independent of the mainshock
magnitude. The specific value of B\aa th's constant $B$
depends only two parameters, the $b$-value of
Gutenberg-Richter distribution and the productivity
exponent $\alpha$.

The second asymptotic for $m>m_c$ corresponds to using
(\ref{31}) to estimate $\langle M_\rho\rangle$, which
yields 
\be 
\Delta_\rho m_2\simeq v_2-\frac{\nu}{w_2}=
\left(1-\frac{\alpha}{b}\right) (\rho-m_0)+ \frac{1}{b}
\left[ \log_{10}\left(\frac{(1-n)b}{n(b-\alpha)}\right) -
\frac{\nu}{\ln 10}\right] ~. 
\label{38} 
\ee 
Note that
$\Delta_\rho m_2$ is increasing with $\rho$ as in
expression (\ref{17}) corresponding to taking only into
account aftershocks of the first generation. This is
natural since the asymptotic for $m>m_c$ corresponds
to $n$ relatively far from $1$ (for a fixed $\rho$),
i.e., far from the critical branching regime, such that
only a few generations play a significant role in the
population of aftershocks. This asymptotic (\ref{38}) is
also identical to the expression (5) of \cite{HSbath}
derived by using the statistical average of the total
number $N_{\rm aft}$ of aftershocks of all generations
triggered by a source of fixed given magnitude. Thus, the
difference between this approximation (\ref{38}) (and
expression (5) of \cite{HSbath}) and the exact expression
(\ref{39}) and its critical universal asymptotic
(\ref{37}) can be traced back to the difference between
the following two kinds of averages: $\langle \ln [N_{\rm
aft}] \rangle$ and $\ln \langle N_{\rm aft} \rangle$.

Figures 7 and 8 show the exact average magnitude
difference (\ref{39}) as a function of the mainshock
magnitude $\rho$ for $b=1$ and different values of the
branching ratio $n$, for $\alpha=0.9$ (Figure 7) and
$\alpha = 0.95$ (Figure 8). As expected from the
condition $m<m_c$ or $n$ closer to $1$ at fixed $\rho$
so that $m_c$ is all the larger according to (\ref{24}),
the largest values of $n$ give almost constant values of
$\Delta_\rho m$ in agreement with the prediction
(\ref{37}). For smaller $n$'s, we can observe a slow
cross-over to the second asymptotic (\ref{38}). By
comparison between Figure 7 and Figure 8, it is clear
that low values of $\alpha$ are not compatible with B\aa
th's law. As confirmed with similar figures obtained for
smaller $\alpha$'s, a value of $\alpha$ at least equal to
$0.9$ seems necessary to obtain a dependence of
$\Delta_\rho m$ roughly independent of $\rho$ over a
large magnitude range. This bound is compatible with some
previous studies
\cite{Felzeretal02,alpha,HelmKaganJackson04} but in
disagreement with others \cite{Consoleetal03,Zhuangetal}.
But, we should remark that such heterogeneity in reported
values of key parameters of the ETAS model such as the
productivity exponent $\alpha$ could be due to the bias
resulting from imperfect account of unobserved seismicity
below the completeness threshold, which may play a
dominant role as explained in \cite{SorWer1,SorWer2}.
Figure 8 also shows that, the closer $n$ is to $1$, the
more independent is $\Delta_\rho m$ with respect to the
mainshock magnitude $\rho$. But, due to inherent
fluctuations in empirical data, $n$ can be as low as
$n=0.8$ for $\alpha \simeq 0.95$ and $\Delta_\rho m$
would still be slowly growing between $1.1$ and $1.3$
over a large magnitude range of $\rho$, so that B\aa th's
law would be approximately verified.

\section{Concluding remarks}

Using the ETAS branching model of triggered seismicity,
we have shown how to calculate exactly the average
difference between the magnitude of a mainshock and the
magnitude of its largest aftershock over all generations.
This average magnitude difference is found empirically to
be independent of the mainshock magnitude and equal to
$1.2$, a universal behavior known as B{\aa}th's law. We
have developed the mathematical formulation in terms of
generating probability functions which allow us to obtain
exact equations and useful approximations to understand
the physical basis for B{\aa}th's law. In particular, we
find that the constancy of the average magnitude
difference (to a value that we term B{\aa}th's constant) is
associated with the critical regime of the ETAS branching
process. Allowing for error bars $\pm 0.1$ for B\aa th's
constant value around $1.2$, our exact analytical
treatment of B\aa th's law provides a new constraint on
two key parameters of the ETAS model, namely the
productivity exponent $\alpha$ and the branching ratio
$n$: $\alpha \gtrsim 0.9$ and $n \gtrsim 0.8$. We have suggested
a novel method for measuring $\alpha$ based
on the predicted renormalization of the Gutenberg-Richter
distribution of the magnitudes of the largest aftershock.
To implement this method, statistical declustering
techniques can be used to identify the spontaneous
sources and their largest aftershocks.
We have also proposed the ``second
B{\aa}th's law for foreshocks'' that  the probability that a main earthquake turns
out to be the foreshock does not depend on its magnitude $\rho$.

{\bf Acknowledgments:} This work is partially supported
by NSF-EAR02-30429, and by the Southern California
Earthquake Center (SCEC) SCEC is funded by NSF
Cooperative Agreement EAR-0106924 and USGS Cooperative
Agreement 02HQAG0008. The SCEC contribution number for
this paper is xxx.

{}

\clearpage

\begin{quote}
\centerline{
\resizebox{12cm}{!}{\includegraphics{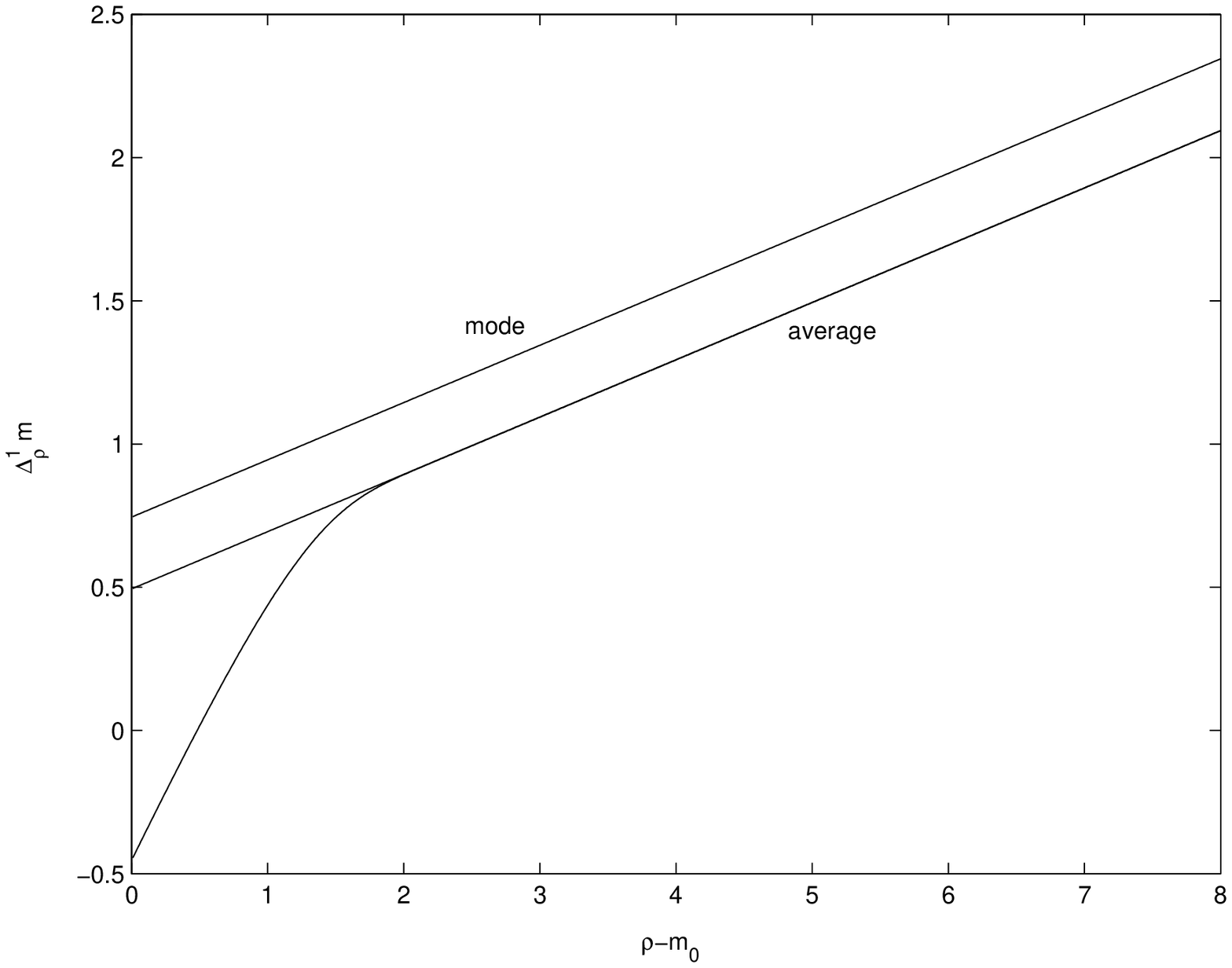}}} {\bf
Fig.~1:} \small{Exact average of the difference $\Delta_1
m= \rho-\langle M_\rho^1\rangle$ obtained using
(\ref{16}) (bottom curve bending down for small $\rho -
m_0$), its large magnitude approximation (\ref{17})
(bottom straight line)  and the difference $\Delta_\rho^1
m_*=v_0(\rho)$ between the mainshock magnitude $\rho$ and
the mode of the magnitude of the largest aftershock among
all aftershocks of the first generation (upper straight
line), for $n=0.9$, $\alpha=0.8$ and $b=1$.}
\end{quote}

\clearpage

\begin{quote}
\centerline{
\resizebox{12cm}{!}{\includegraphics{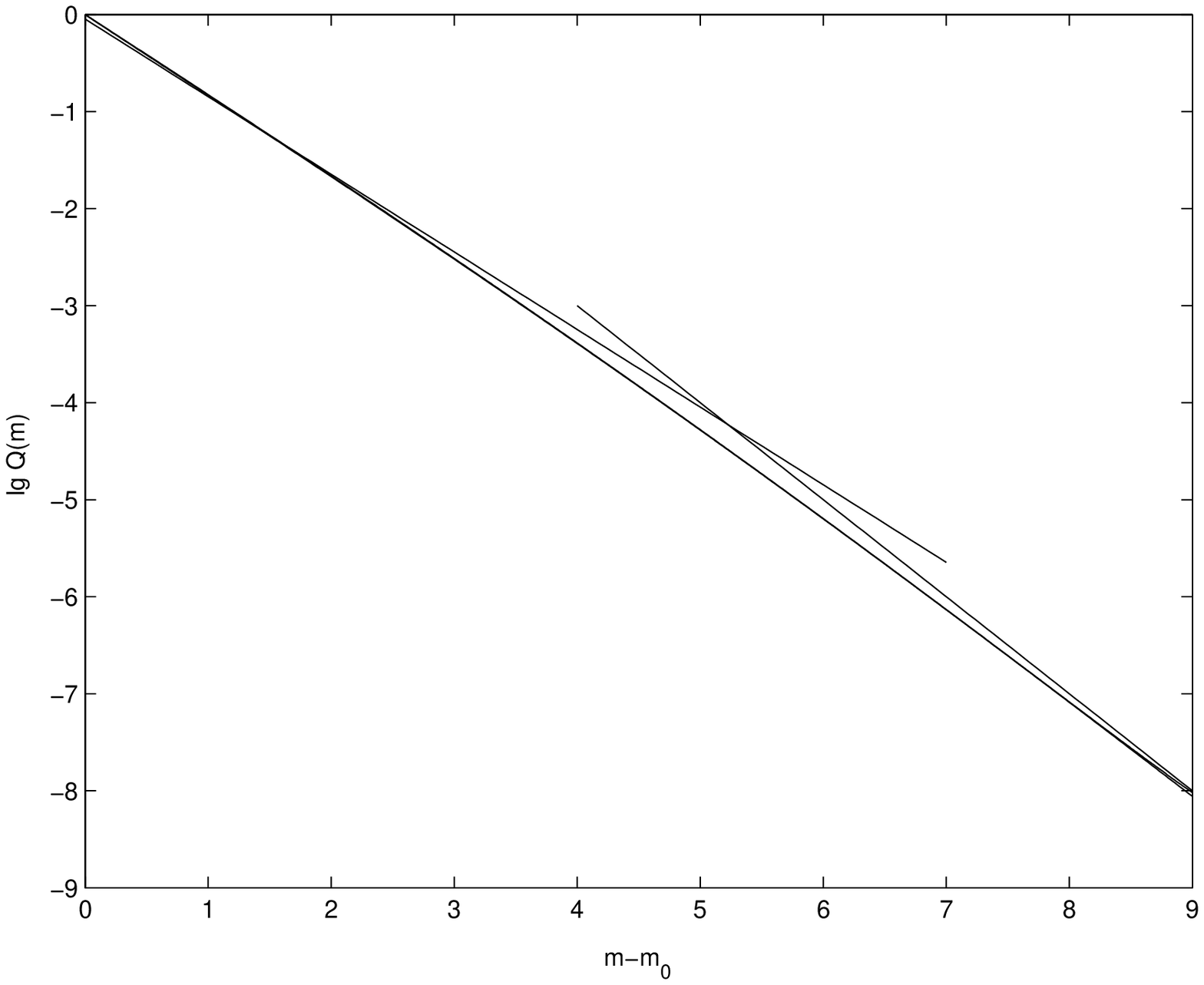}}} {\bf
Fig.~2:} \small{Plot of the decimal logarithm of the
exact CDF $Q(m)$ and its approximation (\ref{23}) (which
actually coincide). Straight lines correspond to the
asymptotics (\ref{25}) and (\ref{26}). }
\end{quote}

\clearpage

\begin{quote}
\centerline{
\resizebox{12cm}{!}{\includegraphics{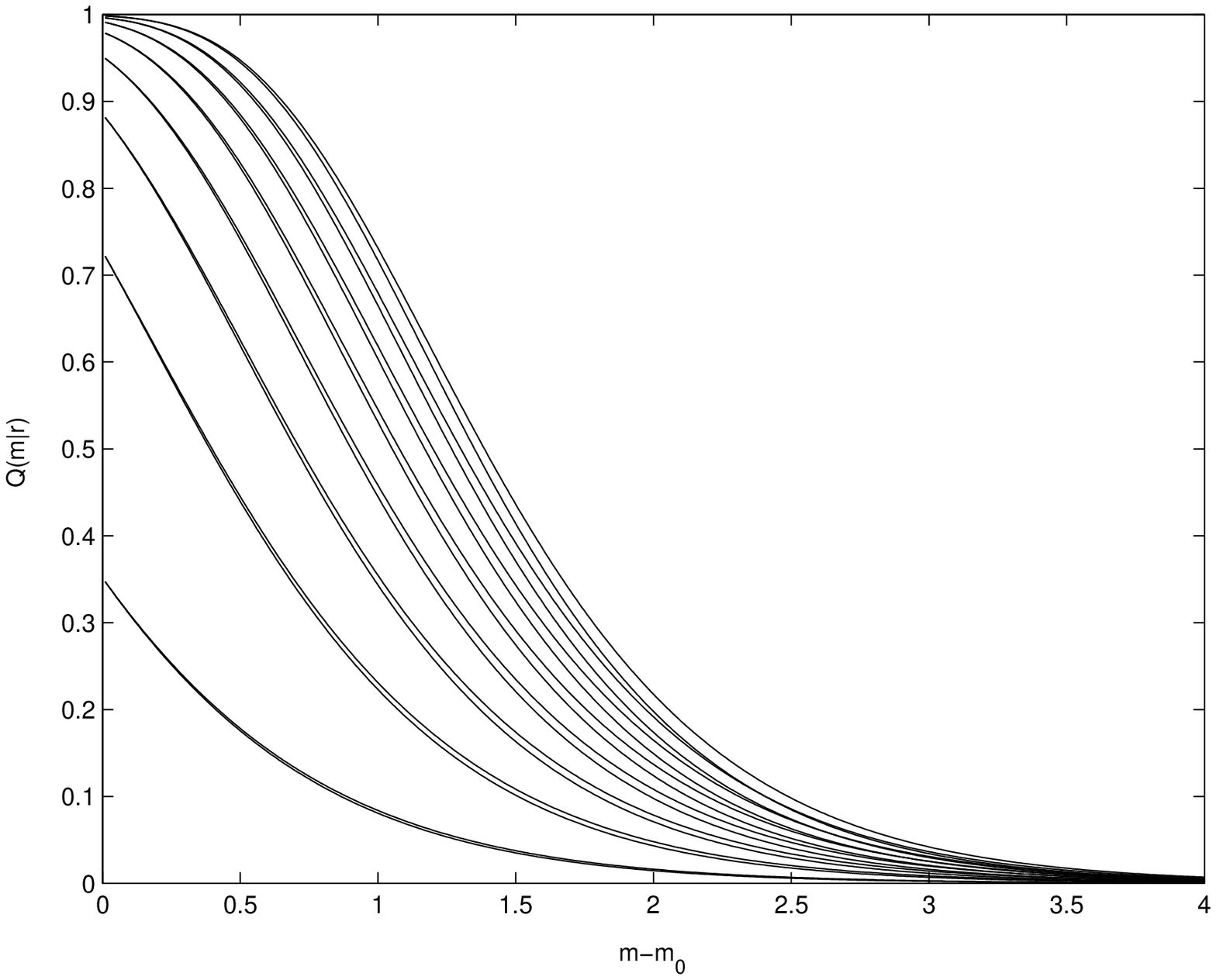}}} {\bf
Fig.~3:} \small{Complementary CDF (\ref{27}) of the
magnitude of the largest event triggered by $r$
spontaneous sources with random magnitudes chosen
according to the Gutenberg-Richter distribution. The
different curves correspond to $r=1;3;5;7;9;11;13;15$
from bottom to top. Each couple of curves corresponds to
$n=0.9$ and $n=1$ respectively.}
\end{quote}

\clearpage

\begin{quote}
\centerline{
\resizebox{12cm}{!}{\includegraphics{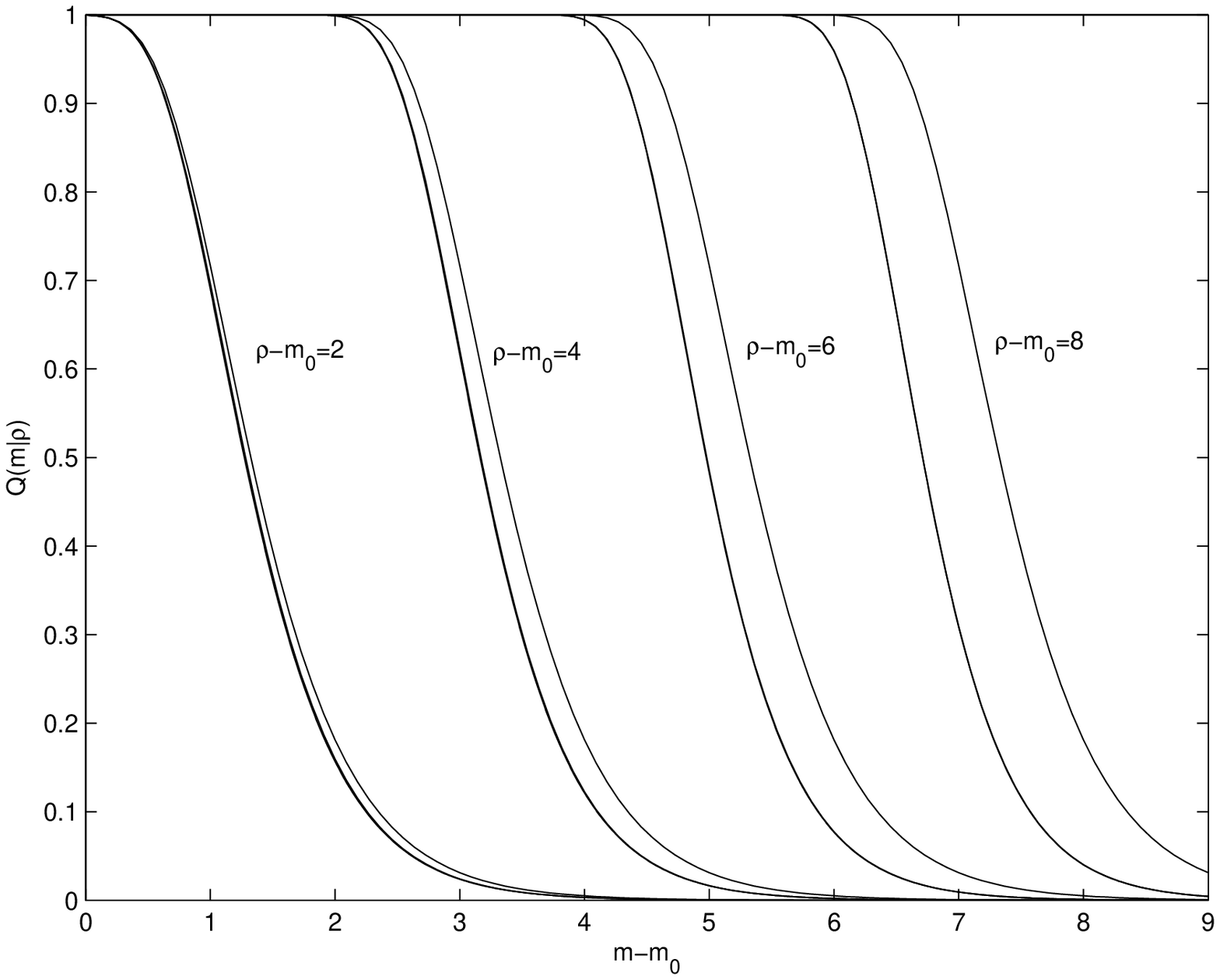}}} {\bf
Fig.~4:} \small{Exact complementary CDF's $Q(m|\rho)$
given by (\ref{18}) (lower curves), their approximations
(\ref{35}) (which are actually undistinguishable from the
exact functions) and the universal approximations
(\ref{29}) (upper curves), for $n=0.9$, $b=1$ and
$\alpha=0.8$ ($\gamma \equiv b/\alpha=1.25$) for four
different values of the spontaneous source magnitude
$\rho$.}
\end{quote}

\clearpage

\begin{quote}
\centerline{
\resizebox{12cm}{!}{\includegraphics{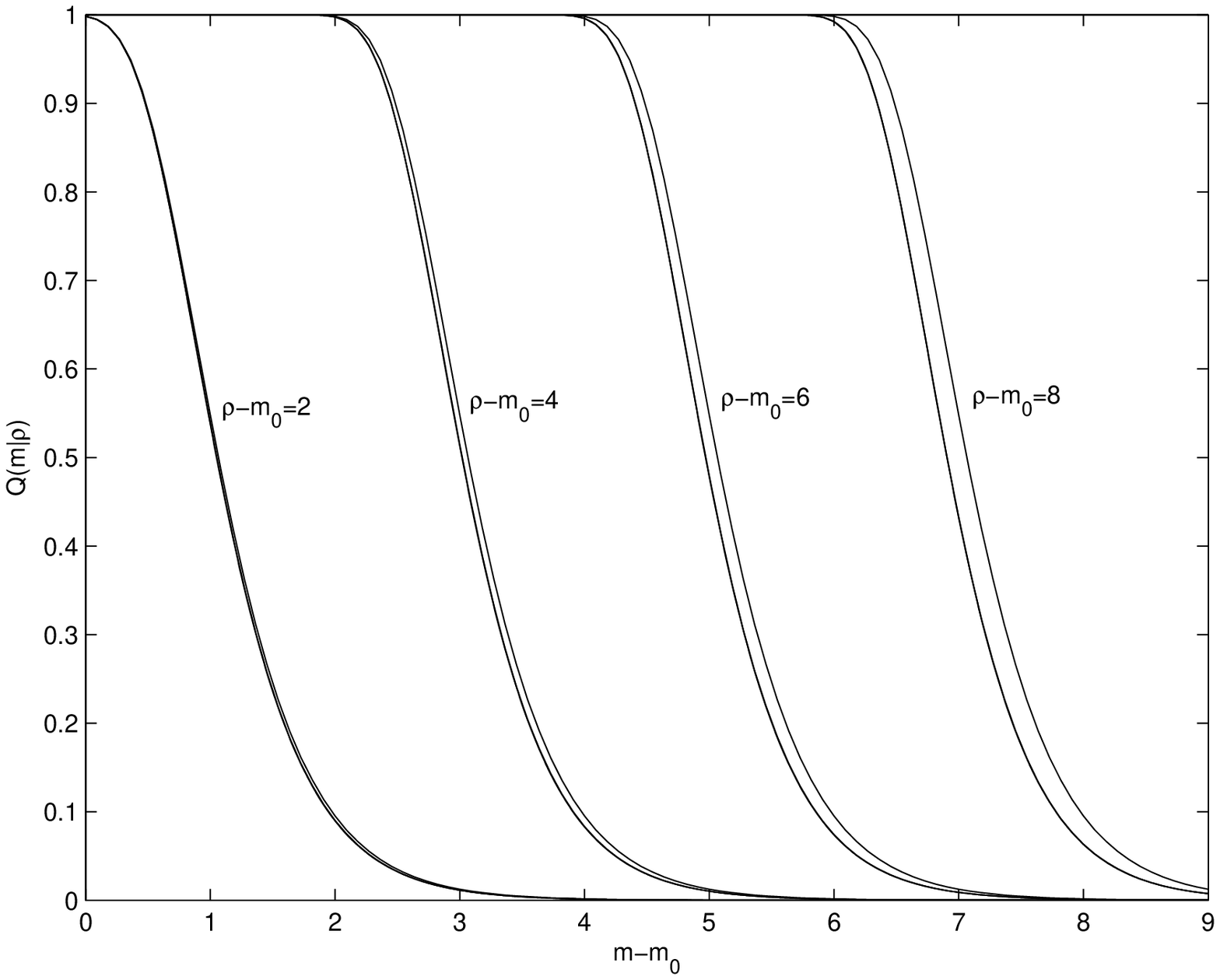}}} {\bf
Fig.~5:} \small{Same as Figure 4 for $\alpha=0.9$
($\gamma=1.11$).}
\end{quote}

\clearpage

\begin{quote}
\centerline{
\resizebox{12cm}{!}{\includegraphics{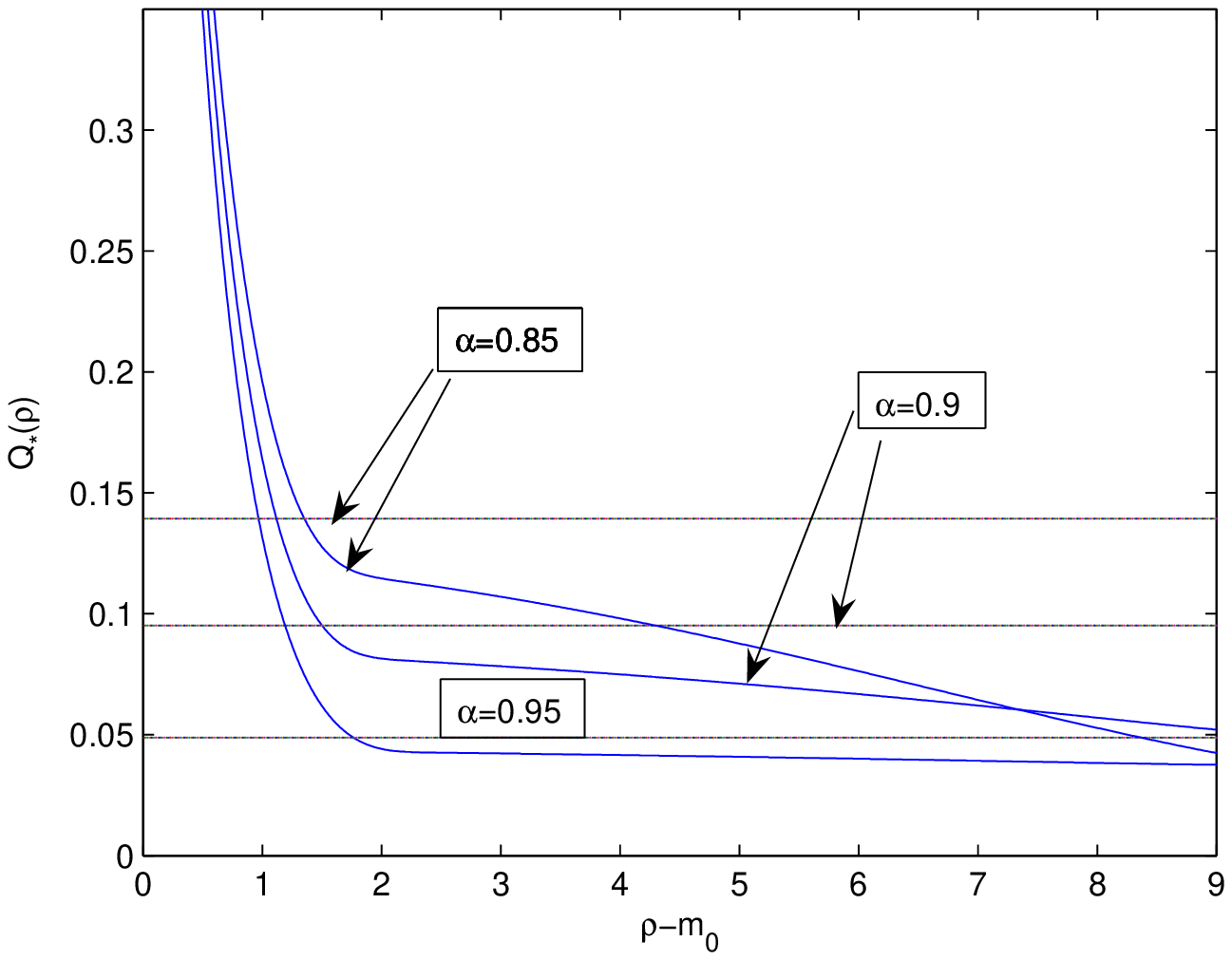}}} 
{\bf Fig.~6:} \small{Plot of $Q_*(\rho)$ as a function of $\rho$,
for $n=0.9$, $b=1$ and different values of $\alpha$. The horizontal
lines correspond to the constants predicted by (\ref{34}).}
\end{quote}
\clearpage

\begin{quote}
\centerline{
\resizebox{12cm}{!}{\includegraphics{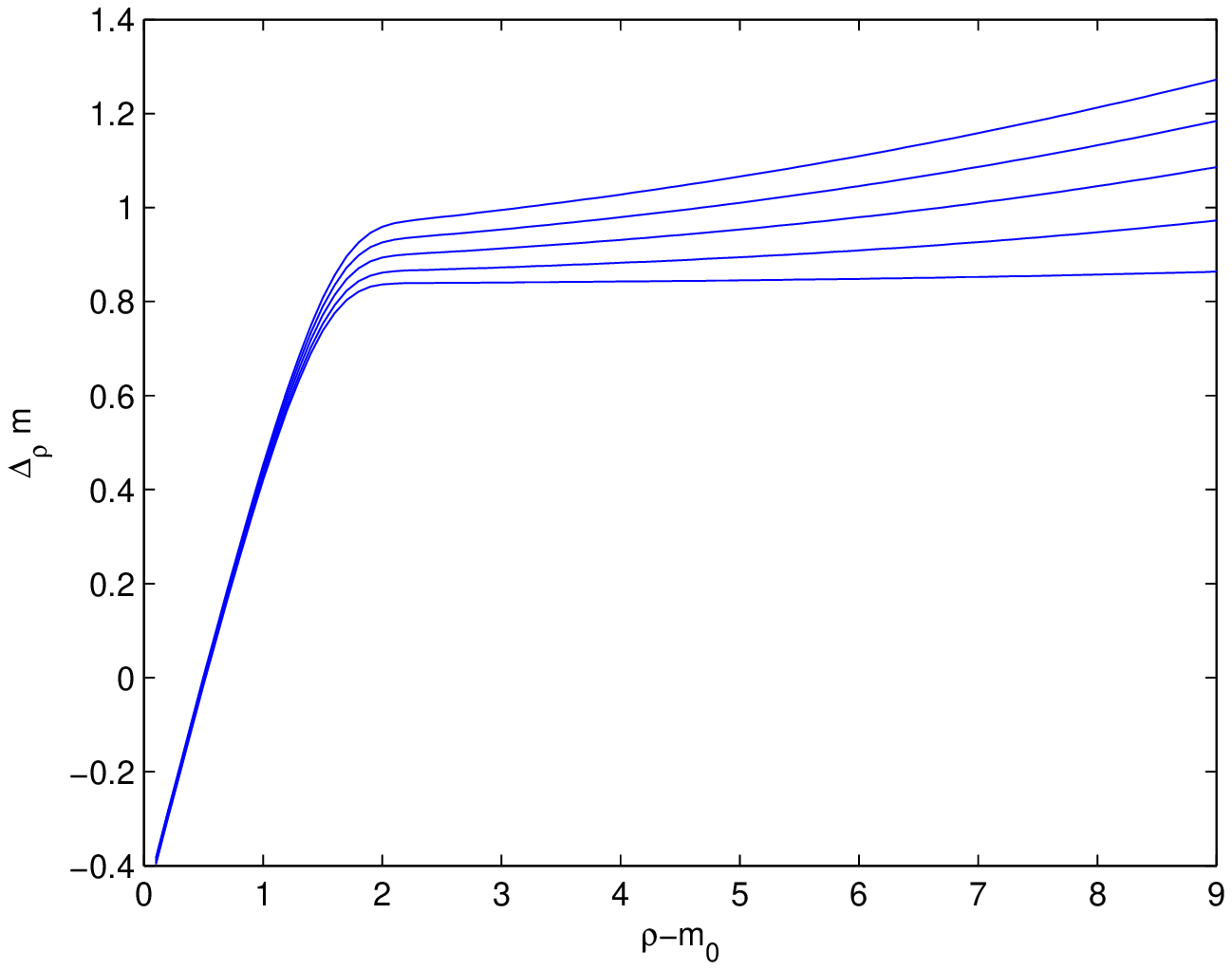}}} {\bf
Fig.~7:} \small{Plot of $\Delta_\rho m$ given by the
exact expression (\ref{39}) for $\alpha=0.9$, $b=1$ and
for different branching ratio. Up to down
$n=0.8;0.85;0.9;0.95; 0.99$. With $n$ tending $n$ to $1$,
the average magnitude difference become closer to the
theoretical B\aa th's constant $B=0.83$. }
\end{quote}

\clearpage

\begin{quote}
\centerline{
\resizebox{12cm}{!}{\includegraphics{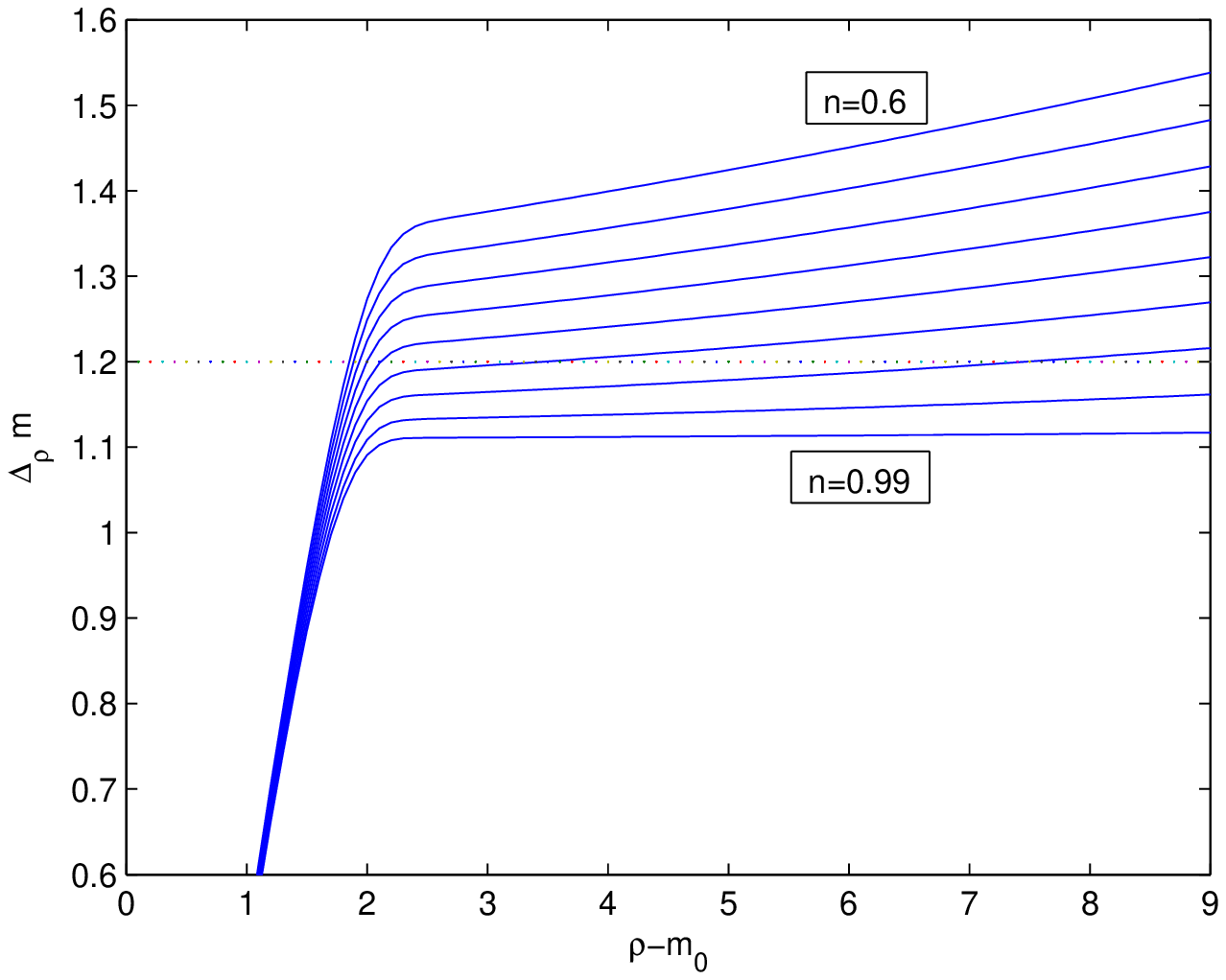}}} {\bf
Fig.~8:} \small{Same as Figure 7 (except for the
magnification) for $\alpha=0.95$ ($\gamma\simeq 1.05$)
giving $B=1.11$.}
\end{quote}

\end{document}